\documentclass[conference]{IEEEtran}
\ifCLASSINFOpdf
\else
\fi
\usepackage{amsfonts,url,graphicx}
\usepackage[cmex10]{amsmath}
\usepackage[tight,footnotesize]{subfigure}
\newcommand{\cv}[1]{\mathbf{#1}}

\newcommand{\her}{^{\rm H}}

\newcommand{\Erw}{\mathbb{E}}

\newcommand{\NT}{N_{\rm T}}
\newcommand{\NR}{N_{\rm R}}

\renewcommand{\v}[1]{\underline{\mathbf{#1}}}

\begin{document}
%
\title{Performance of Interference Alignment in Clustered \\ Wireless Ad Hoc Networks}


\author{\IEEEauthorblockN{Roland Tresch, Maxime Guillaud}
\IEEEauthorblockA{FTW (Telecommunications Research Center Vienna)\\
Vienna, Austria\\
e-mail: \tt{\{tresch,guillaud\}@ftw.at}
}
}


%


\maketitle

\begin{abstract}
Spatial interference alignment among a finite number of users is proposed as a technique to increase the probability of successful transmission in an interference limited clustered wireless ad hoc network. Using techniques from stochastic geometry, we build on the work of Ganti and Haenggi dealing with Poisson cluster processes with a fixed number of cluster points and provide a numerically integrable expression for the outage probability using an intra-cluster interference alignment strategy with multiplexing gain one. For a special network setting we derive a closed-form upper bound. We demonstrate significant performance gains compared to single-antenna systems without local cooperation.

\end{abstract}


%
\IEEEpeerreviewmaketitle

\section{Introduction}
\label{section_introduction}

The fundamental performance limits of wireless ad hoc networks were studied in the seminal paper \cite{OezguerLevequeTseit07}. Short range communication in local clusters in conjunction with distributed multiple-input multiple-output (MIMO) communication is shown to provide optimal capacity scaling. The exact characterization of the interference including fading effects and more general node distributions as in \cite{OezguerLevequeTseit07} is a challenging problem. Analytically tractable results on the interference, outage probability and transmission capacity of large random wireless networks whose underlying node distribution is a Poisson clustered process were established in \cite{GantiHaenggiit09}. Techniques from stochastic geometry \cite{StochasticGeometry_Stoyan}, \cite{HaenggiStochGeom09} were used to characterize the distribution of interference from other concurrent transmissions at a reference receiving node as a function of the density of the transmitters, the path-loss exponent and the fading distribution.

Interference alignment was first considered in \cite{maddahali_isit06} as a coding technique for the two-user MIMO X channel, where it was shown to achieve multiplexing gains strictly higher than that of the embedded MIMO interference channel, multiple-access channel, and broadcast channel taken separately. While requiring perfect channel knowledge, this coding technique is based only on linear precoding at the transmitters and interference suppression filtering at the receivers. This transmission technique was later generalized to the $K$-user interference channel \cite{Cadambeit08}, where it was shown to achieve almost surely a sum-rate multiplexing gain of $\frac{K}{2}$ per time and frequency dimension. In comparison, independent operation of $K$  \emph{isolated} point-to-point links would incur a sum-rate multiplexing gain of $K$ per dimension. Thanks to the alignment of all interfering signals in the same subspace from the point of view of all receivers simultaneously, interference can be removed simply through interference suppression filtering.

In \cite{Cadambeit08}, an explicit formulation of the precoding vectors achieving interference alignment is presented for single-antenna nodes with time-varying channels. In the multiple-antenna case, no such closed-form solution is known, although achievability results on multiplexing gains are available \cite{Gou08}, \cite{cadambe-2009}. An iterative algorithm was introduced in \cite{Gomadamit08} to find numerically the precoding matrices achieving interference alignment. Feasibility results for the \emph{constant} MIMO interference channel are presented in \cite{Cadambeit08}, \cite{TreschGuillaudRiegler_SSP09}, and \cite{yetis_gou_09}. Extending alignment-based interference suppression schemes to larger networks, \cite{Johnson_09} applies interference alignment to large scale Gaussian interference networks and derives bounds on the sum capacity. Fading effects were omitted.

The goal of this contribution is to evaluate the performance of MIMO interference alignment in large clustered wireless networks. In order to decrease feedback signaling and the number of antennas that need to be deployed per node, we propose clustered interference alignment among a finite number of cooperating users, while the rest of the network contributes non-aligned interference. We consider an intra-cluster interference alignment scheme with multiplexing gain one per transmitter-receiver pair and quantify the benefit of suppressing interfering nodes in the vicinity of the intended transmitter in terms of outage probability.

This article is organized as follows: the system model, the principle of intra-cluster interference alignment, and feasible cluster settings are introduced in Section \ref{section_sysmodel}. Section \ref{section_success_prob} derives expressions for the outage probability, while Section \ref{section_simres} presents simulation results and performance comparisons.

\section{System Model}
\label{section_sysmodel}

\begin{figure}[h]
  \centering
  \includegraphics[width=7cm]{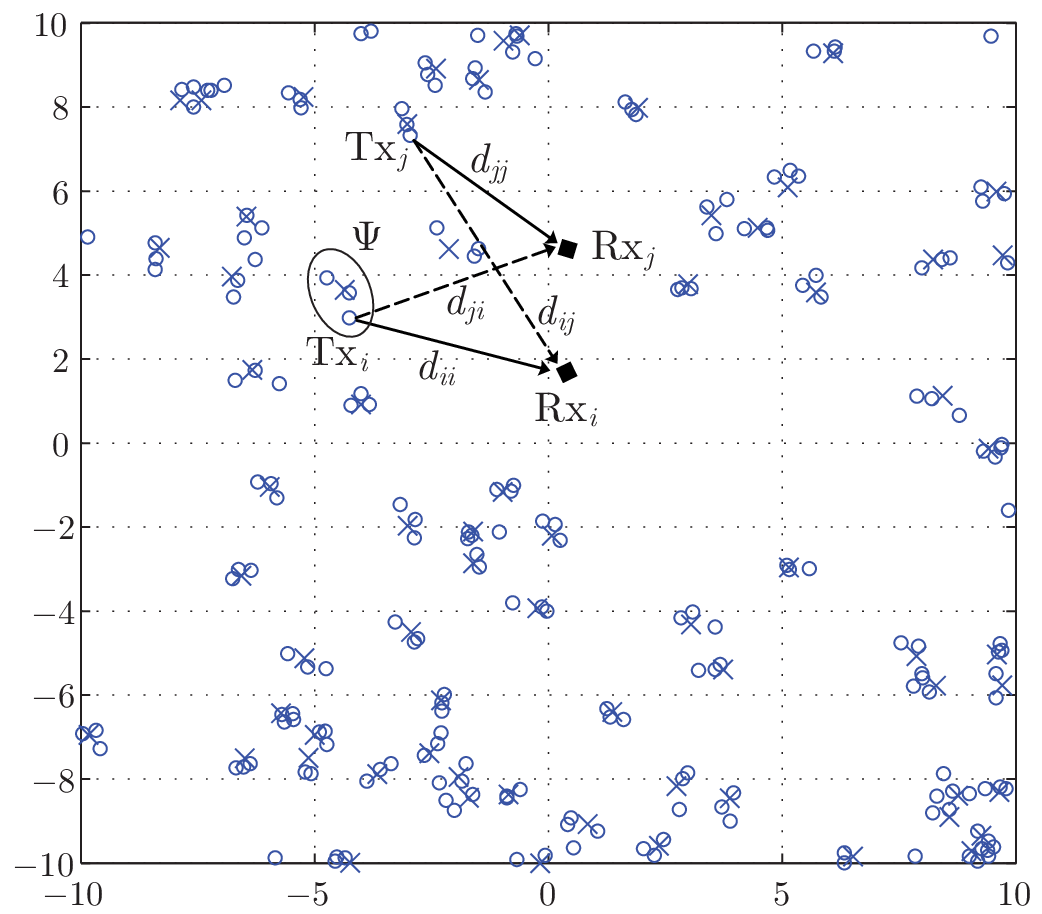}
  \caption{Sample of the considered Poisson cluster process $\lambda_p=0.2$, a fixed number of cluster points $\bar{c}=3$ and scattering parameter $\sigma=0.25$. Parent points are depicted by crosses and daughter points by circles}
  \label{Fig:sample_cluster_process}
\end{figure}

Let us introduce the model for the clustered wireless ad hoc network. The clustering of nodes may be due to geographical factors, hot spots with a high user density or induced by the channel access scheme. The location of the transmitting nodes is modeled as a Neyman-Scott cluster process which is a stationary and isotropic Poisson cluster process $\Phi$ on the infinite plane $\mathbb{R}^2$ \cite{StochasticGeometry_Stoyan}, an example of which is depicted in Fig. \ref{Fig:sample_cluster_process}. A Neyman-Scott process results from homogeneous independent clustering applied to a stationary Poisson process \cite{GantiHaenggiit09}. The cluster process consists of a parent Poisson process of density $\lambda_p$. The parent points themselves (depicted by crosses in Fig. \ref{Fig:sample_cluster_process}) are not included but serve as reference points for the daughter points that are scattered independently and with identical distribution around the position of the parent points. The scattering density function $f(x)$ is chosen as
\begin{align}\label{equ_scatter}
  f(x) = \frac{1}{2\pi\sigma^2}\exp\left(-\frac{\|x\|^2}{2\sigma^2}\right),
\end{align}
with the vector $x$ containing the two dimensional coordinates relative to the parent point. Hence, each point is scattered using a circularly symmetric normal distribution with variance $2\sigma^2$. Since the scattering density of the representative cluster is isotropic, the whole cluster process $\Phi$ is isotropic. Here, we assume that the number of points $\bar{c}$ in a representative cluster $\Psi$ is fixed. The clusters form a partition of $\Phi$. The overall intensity of the cluster process is therefore the product of the parent process intensity and the number of nodes per cluster $\lambda=\lambda_p\bar{c}$.

We can write the point process as a random set $\Phi=\{x_1,x_2,\ldots,x_\infty\}$, where $x_j$ is the coordinate of the point in $\mathbb{R}^2$ \cite{HaenggiStochGeom09}. Let us denote the set of indices of the points $\phi=\{1,2, \ldots,\infty\}$ where $\psi=\{k,k+1,\ldots,k+\bar{c}-1\}$ is a subset of $\phi$ containing the indices of $\Psi$. Each transmitter $j\in\phi$ is assumed to transmit at unit power. The receivers, depicted by squares in Fig. \ref{Fig:sample_cluster_process}, are not considered a part of the process. That is, the model excludes the partner selection problem and focuses on the notion of a common distance that information travels in the network \cite{GantiHaenggiit09}. Every transmitter and receiver is equipped with $\NT$ and $\NR$ antennas, respectively. The transmitters inside of a cluster $\Psi$ use a spatial interference alignment (IA) scheme \cite{Tresch_Guillaud_PIMRC09} in order to suppress the interference that they cause to each other, i.e. the members of the same cluster $\Psi$ cooperate and adjust the spatial structure of their transmitted signals in order to avoid interfering with each other. The rest of the network $\Phi\setminus\Psi$ contributes non-aligned interference. While not optimal in general, treating the inter-cluster interference as noise is, in fact, optimal in the Gaussian weak interference regime \cite{ShangKramerChenit09}. In particular, this case arises when the cluster density $\lambda_p$ is small.

Let us focus on a receiver $i\in \phi$, see Fig. \ref{Fig:sample_cluster_process}. The discrete-time signal received at a given time instant is the superposition of the signals transmitted by the $\bar{c}$ transmitters of the considered cluster $\Psi$, and the transmitters of the rest of the network $\Phi\setminus\Psi$, weighted by their respective channel gains and path-loss coefficients. Namely, the signal at receiver $i$ can be written as
\begin{align}
        \v{y}_i =  \sum_{j\in \phi} \sqrt{\gamma_{ij}}\cv{H}_{ij}\v{v}_{j} s_j + \v{n}_i, \label{intfchannel}
\end{align}
where $s_j\in \mathbb{C}$ represents the scalar signal transmitted by node $j$, and $\v{v}_j\in \mathbb{C}^{\NT\times 1}$ is the associated precoding vector which will be further specified in Subsection \ref{subsection_Feasible}. $[\cv{H}_{ij}]_{i,j\in\phi}$ are complex $\NR \times \NT$ matrices representing the MIMO channels between transmitter $j$ and receiver $i$. $\gamma_{ij}=\sqrt{g(d_{ij})}$ is the path-loss model on the same link with $d_{ij}$ the distance between transmitter $j$ and receiver $i$ and $g(\cdot)$ the path-loss function. We assume a flat-fading channel model.
$\v{n}_i$ is a noise term, accounting for the thermal noise generated in the radio frequency front-end of the receiver and interference from sources other than the considered transmitters $j\in \phi$. 

\subsection{Intra-Cluster IA and Feasible Cluster Settings}\label{subsection_Feasible}

Our aim is to suppress the intra-cluster interference inside of each cluster $\Psi$ individually, by means of spatial IA. More specifically, we use IA with multiplexing gain one for each of the $\bar{c}$ transmitter-receiver pairs. This strategy is suboptimal in general, but chosen for analytical tractability. Therefore, the transmitters $i\in\psi$ share their knowledge about the frequency-flat channel matrices $[\cv{H}_{ij}]_{i,j\in\psi}$ and choose a precoding vector $\v{v}_{i}$ in order to steer their transmitted signal into a subspace of minimum dimension at each unintended receiver, where it creates interference. The receiver uses a projection vector $\v{u}_i$ on the subspace orthogonal to the intra-cluster interference in order to suppress unintended signals from transmitters $(j\neq i) \in\psi$. The remaining dimension of the received signal space is used for interference free communication with the intended transmitter.

Therefore, for every cluster $\Psi$, IA is achieved in the spatial domain with degree of freedom one iff there exists $\NT\times 1$ unit-norm vectors (precoding vectors) $\v{v}_{i}$ and $\NR\times 1$ unit-norm vectors (interference suppression vectors) $\v{u}_{i}$ such that, for all $i\in\psi$,
\begin{align}
\v{u}_{i}\her \cv{H}_{ij}\v{v}_{j} &= 0,\forall (j\neq i) \in\psi, \quad \mathrm{and} \label{equ_IAcondA}\\
\textrm{rank}\left(\v{u}_{i}\her \cv{H}_{ii}\v{v}_{i} \right)&= 1. \label{equ_IAcondB}
\end{align}
Under the assumption that the channel coefficients $\cv{H}_{ij}$ are drawn independently and identically from a continuous distribution, it appears from \cite{TreschGuillaudRiegler_SSP09}, \cite{yetis_gou_09} that the existence (with probability one) of an IA solution depends solely on the dimensions of the problem ($\bar{c},\NT,\NR)$, and not on the particular channel realization. In the particular case of multiplexing gain one, solutions exist for the $\bar{c}$-user interference channel iff
\begin{align}\label{equ_IA_conj}
\NR+\NT-1 \geq \bar{c}.
\end{align}
The iterative algorithm \cite[Algorithm 1]{Gomadamit08} or for symmetric MIMO systems the closed-form solution in \cite{TreschGuillaudRiegler_SSP09} can be used to find the precoding and interference suppression vectors.

\subsection{Analysis of the Equivalent Channel}

Let us now analyze the input-output relation of the system after interference suppression inside the clusters. Projecting the receive signal $\v{y}_i$ (\ref{intfchannel}) onto the orthogonal subspace of the interference, using $\v{u}_{i}$, suppresses all intra-cluster interference. Thus, the signal after projection at receiver $i$ yields
\begin{align}\label{equ_ybar_global}
\bar{y}_i = \v{u}_{i}\her \v{y}_i = & \sqrt{\gamma_{ii}}\v{u}_{i}\her\cv{H}_{ii}\v{v}_{i} s_i + \nonumber \\
    & \sum_{j\in (\phi\setminus\psi)} \sqrt{\gamma_{ij}}\v{u}_{i}\her\cv{H}_{ij} \v{v}_{j} s_{j} + \v{u}_{i}\her\v{n}_i.
\end{align}
Here, we used the fact that the interference from transmitters within $\Psi$ is perfectly suppressed, due to \eqref{equ_IAcondA}. The effective channel $\bar{h}_{ii}=\v{u}_{i}\her\cv{H}_{ii}\v{v}_{i}$ as seen by receiver $i$ is scalar. Furthermore, since $\v{u}_{i}\her\v{u}_{i}=\v{v}_{i}\her\v{v}_{i}=1$, $\bar{h}_{ii}$ is a scalar coefficient with the same variance as the components in $\cv{H}_{ii}$. Similarly, the effective noise term $\bar{n}_i=\v{u}_i\her \v{n}_i$, is a scalar Gaussian coefficient with the same variance $\sigma_n^2$ as the noise vector $\v{n}_i$ and $\bar{h}_{ij}=\v{u}_{i}\her\cv{H}_{ij}\v{v}_{j}$.

\begin{figure}
  \centering
  \subfigure[feasible MIMO setting for intra-cluster IA]{\label{Fig:effective_channel_a} \includegraphics[width=4cm]{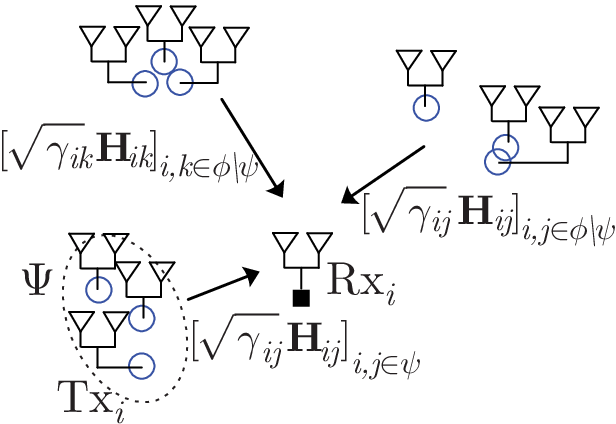}}\quad
  \subfigure[equivalent SISO system (\ref{equ_ybar_equivalent}) after interference suppression filtering]{\label{Fig:effective_channel_b} \includegraphics[width=4cm]{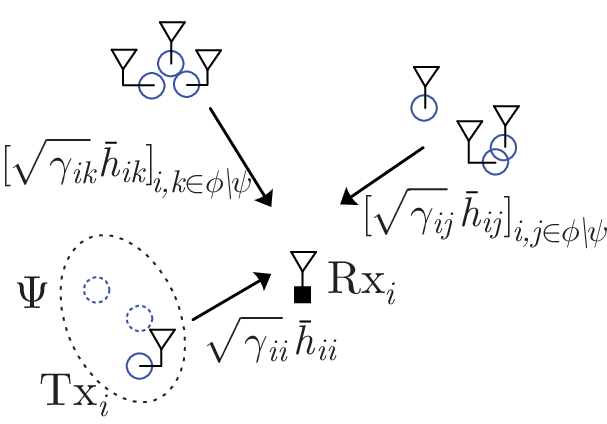}}
  \caption{Cluster setting with $\bar{c}=3$ cluster points, $2\times2$ MIMO, and equivalent input-output relation}
  \label{Fig:effective_channel}
\end{figure}

Thus, an equivalent scalar input-output relation of the system after interference suppression yields
\begin{align}\label{equ_ybar_equivalent}
\bar{y}_i = & \sqrt{\gamma_{ii}}\bar{h}_{ii}s_i +\sum_{j\in (\phi\setminus\psi)} \sqrt{\gamma_{ij}}\bar{h}_{ij} s_{j} + \bar{n}_i,
\end{align}
which can be interpreted as a system with single antenna terminals where the intra-cluster interference has been completely suppressed, as pictured in Fig. \ref{Fig:effective_channel}.

\section{Transmission Success Probability}
\label{section_success_prob}

Let the desired transmitter be located at the origin and the receiver at location $z$ at distance $d_{ii}=\|z\|$ from the transmitter \cite{GantiHaenggiit09}. Communication is successful, if the signal-to-interference-plus-noise ratio (SINR) exceeds a certain threshold $T$ that depends on the physical layer parameters such as rate of transmission, modulation and coding. Thus, the probability of success for this pair is given by \cite[eq. (31)]{GantiHaenggiit09}
\begin{align}\label{equ_Pr_succ}
  \mathbb{P}(\textrm{success})=\mathbb{P}\left(\frac{|\bar{h}_{ii}|^2 \gamma_{ii}}{\sigma_n^2+I_{\phi\setminus\psi}}\geq T\right),
\end{align}
with $I_{\phi\setminus\psi}=\sum_{j\in (\phi\setminus\psi)}|\bar{h}_{ij}|^2 \gamma_{ij}$ the accumulated inter-cluster interference from the rest of the network. We now assume Rayleigh fading, i.e. the received power $|\bar{h}_{ii}|^2$ is exponentially distributed with mean $\mu$. Using tools from stochastic geometry, in \cite[Appendix II]{GantiHaenggiit09} it is shown that the success probability (when we neglect the noise $\sigma_n^2=0$\footnote{In a dense ad hoc network, the noise is typically negligible and SIR can be used interchangeably with SINR without any loss of accuracy.}) is given by
\begin{align}\label{equ_Pr_succ_fixed}
  \mathbb{P}(\textrm{success})=p_s=\exp\bigg\{-\lambda_p\underbrace{\int_{\mathbb{R}^2}1-\tilde{\beta}(z,y)^{\bar{c}}\textrm{d}y}_{\xi(z)}\bigg\},
\end{align}
where
\begin{align}\label{equ_beta_tilde}
  \tilde{\beta}(z,y)=\int_{\mathbb{R}^2}\frac{f(x)}{1+\frac{Tg(x-y-z)}{g(z)}}\textrm{d}x,
\end{align}
and independent of the mean $\mu$. We denote the two dimensional integral in (\ref{equ_Pr_succ_fixed}) by $\xi(z)$. The integral over $y$ in $(\ref{equ_Pr_succ_fixed})$ originates from the Campbell-Mecke theorem applied to the homogeneous Poisson parent process. The expression $\tilde{\beta}(z,y)^{\bar{c}}$ via the integral over $x$ in (\ref{equ_beta_tilde}) accounts for the probability generating functional of Poisson cluster processes with a fixed number of cluster points. The expression for $\mathbb{P}(\textrm{success})$ in \cite[Appendix II]{GantiHaenggiit09} accounts for intra-cluster interference via the term $\int\tilde{\beta}(z,y)^{\bar{c}-1}f(y)\textrm{d}y$. Since we suppress the intra-cluster interference, see Fig. \ref{Fig:effective_channel_b}, this term was neglected in (\ref{equ_Pr_succ_fixed}).

\subsection{Bounds on the Transmission Success Probability}

In this subsection, we fix the path-loss model $g(d_{ij})=d_{ij}^{-\alpha}$ with $2 \leq\alpha\leq 4$ and seek for bounds on the transmission success probability. Thus, our attempt is to simplify (\ref{equ_Pr_succ_fixed}).

We tackle the upper bound on $\mathbb{P}(\textrm{success})$. Therefore, we are looking for an upper bound on $\tilde{\beta}(z,y)$ (\ref{equ_beta_tilde}). If one considers $x$ in (\ref{equ_beta_tilde}) as a random vector $X$ with two-dimensional density function $f(\cdot)$, we can write
\begin{align}
  \tilde{\beta}(z,y) &=\Erw\left[\frac{1}{1+\frac{T\|X-y-z\|^{-\alpha}}{\|z\|^{-\alpha}}}\right]=\Erw\left[\frac{1}{1+\frac{T (U^2)^{-\alpha/4}}{\|z\|^{-\alpha}}}\right],
\end{align}
where the second equality comes from a change of variable $U=\|X-y-z\|^{2}$. Since $X$ is assumed to have a symmetric normal distribution with variance $2\sigma^2$, the random variable $U$ has a non-central chi-square distribution with mean $2\sigma^2+\|y+z\|^{2}$ and variance $4\sigma^4+4\sigma^2\|y+z\|^{2}$. Furthermore, since $1/(1+T V^{-\alpha/4}/\|z\|^{-\alpha})$ is concave for $V>0$ and $\alpha\leq 4$, we apply Jensen's inequality to get
\begin{align}\label{equ_jensons}
  \tilde{\beta}(z,y) &= \Erw\left[\frac{1}{1+\frac{T (U^2)^{-\alpha/4}}{\|z\|^{-\alpha}}}\right]  \leq \frac{1}{1+\frac{T\Erw[U^2]^{-\alpha/4}}{\|z\|^{-\alpha}}}\nonumber \\ &=\frac{1}{1+\frac{T\left((2\sigma^2+\|y+z\|^{2})^2+4\sigma^4+4\sigma^2\|y+z\|^{2}\right)^{-\alpha/4}}{\|z\|^{-\alpha}}},
\end{align}
where we used the fact that the second raw moment of $U$ is $(2\sigma^2+\|y+z\|^{2})^2+4\sigma^4+4\sigma^2\|y+z\|^{2}$. Since we found an upper bound on $\tilde{\beta}(z,y)$, we lower bound $\xi(z)$ in (\ref{equ_Pr_succ_fixed}) with (\ref{equ_jensons}) as
\begin{align}\label{equ_int_LB}
& \xi(z)\geq \nonumber \\
& \int_{\mathbb{R}^2}1-\left(\frac{1}{1+\frac{T\left((2\sigma^2+\|y+z\|^{2})^2+4\sigma^4+4\sigma^2\|y+z\|^{2}\right)^{-\alpha/4}}{\|z\|^{-\alpha}}}\right)^{\bar{c}}\textrm{d}y.
\end{align}
Shifting the integrand in (\ref{equ_int_LB}) towards direction $-z$ and change of variables leads to an equivalent expression for the lower bound on $\xi(z)$, i.e.
\begin{align}\label{equ_int_LB2}
& \int_{\mathbb{R}^2}1-\left(\frac{1}{1+\frac{T\left((2\sigma^2+\|y\|^{2})^2+4\sigma^4+4\sigma^2\|y\|^{2}\right)^{-\alpha/4}}{\|z\|^{-\alpha}}}\right)^{\bar{c}}\textrm{d}y \nonumber \\
= & 2\pi\int_0^{\infty}\left(1-\left(\frac{1}{1+\frac{T\left((2\sigma^2+r^{2})^2+4\sigma^4+4\sigma^2 r^{2}\right)^{-\alpha/4}}{\|z\|^{-\alpha}}}\right)^{\bar{c}}\right)r\textrm{d}r \nonumber \\
= & \pi \int_{4\sigma^2}^{\infty}1-\left(\frac{1}{1+\frac{T (s^2-8\sigma^4)^{-\alpha/4}}{\|z\|^{-\alpha}}}\right)^{\bar{c}}\textrm{d}s,
\end{align}
where the first equality comes from a change of the cartesian coordinates $y=(y_1,y_2)$ into polar coordinates, i.e. $y_1=r\cos\varphi$ and $y_2=r\sin\varphi$. The second equality follows from change of variable $s=r^2+4\sigma^2$. Thus, the upper bound on the transmission probability writes
\begin{align}\label{equ_UB_1dim_int}
  p_s\leq \exp\left\{-\lambda_p \pi \int_{4\sigma^2}^{\infty}1-\left(\frac{1}{1+\frac{T (s^2-8\sigma^4)^{-\alpha/4}}{d_{ii}^{-\alpha}}}\right)^{\bar{c}}\textrm{d}s \right\},
\end{align}
where we replaced $\|z\|$ by $d_{ii}$. Hence, the upper bound for probability of success depends only on the distance between transmitter $i$ and receiver $i$ and is irrespective of the relative position $z$.

\subsubsection{Closed-form solution for a special case}

If we consider $\alpha=4$ and $\sigma^2\ll 1$, i.e. the cluster size is typically small, we can find a closed-form upper bound for $p_s$. Neglecting $8\sigma^4$ in the integrand of (\ref{equ_int_LB2}) and changing of variables $t=1/s$ leads to
\begin{align}
   \xi(z) &\geq \pi \int_{0}^{1/(4\sigma^2)} \sum_{k=1}^{\bar{c}} \frac{\frac{T}{d_{ii}^{-\alpha}}}{\left(\frac{T}{d_{ii}^{-\alpha}} t^2+1\right)^k} \textrm{d}t \nonumber\\
   &= \pi \left( \delta(\bar{c}) d_{ii}^2\sqrt{T}\tan^{-1}\left(\frac{d_{ii}^2 \sqrt{T}}{4\sigma^2}\right)+\mathcal{R}(d_{ii},T,\sigma^2)\right),
\end{align}
with $\delta(\bar{c})=\sum_{k=0}^{\bar{c}-1}(-1)^k \left( \begin{array}{c} -1/2 \\ k \end{array}\right)$ and a residual term $\mathcal{R}(d_{ii},T,\sigma^2)$. Neglecting the term $\mathcal{R}(d_{ii},T,\sigma^2)$ leads to
\begin{align}\label{equ_UB_closed_form}
  p_s\leq \exp\left\{-\lambda_p \pi \delta(\bar{c}) d_{ii}^2\sqrt{T}\tan^{-1}\left(\frac{d_{ii}^2 \sqrt{T}}{4\sigma^2}\right) \right\}.
\end{align}

\section{Simulation Results}
\label{section_simres}

In this section, we present simulation results for different settings of the Poisson cluster process and plot the closed-form upper bound. The curves for the probability of success were derived by numerical integration of (\ref{equ_Pr_succ_fixed}) and verified by Monte-Carlo simulation of (\ref{equ_Pr_succ}) over different transmitter positions and fading coefficients. The standard power law $g(d_{ij})=d_{ij}^{-\alpha}$ with $\alpha=4$ and a threshold $T=0.1$ is used.

\begin{figure}[h]
  \centering
  \includegraphics[width=7.5cm]{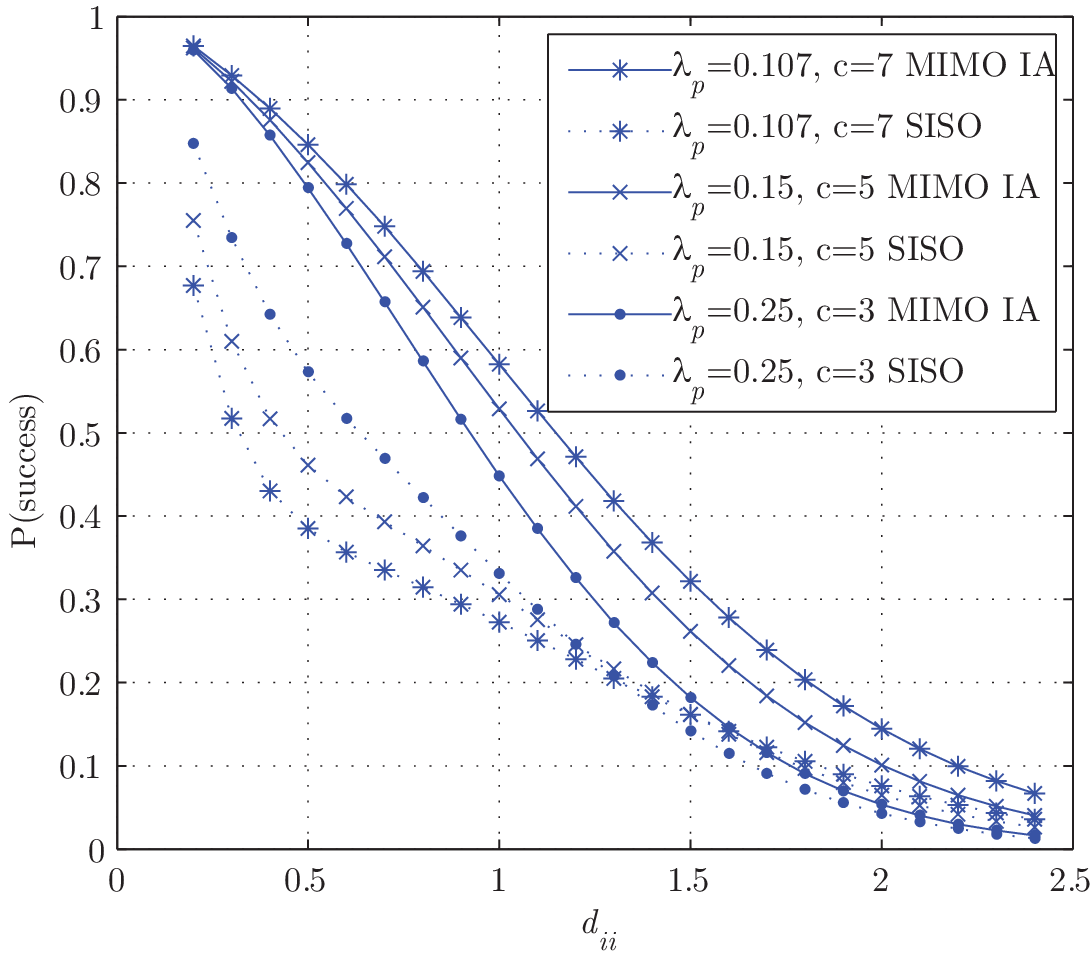}
  \caption{$\mathbb{P}(\textrm{success})$ versus $d_{ii}$ for $\lambda=\lambda_p\bar{c}$ fixed, $\sigma$=0.25, $T=0.1$, and $\alpha=4$, comparison of SISO clustered network with MIMO intra-cluster IA}
  \label{Fig:Res3}
\end{figure}

In Fig. \ref{Fig:Res3} the probability of success of the intra-cluster IA ("MIMO IA") and the corresponding single-antenna network ("SISO") with the same underlying node distribution suffering from intra- and inter-cluster interference is plotted versus $d_{ii}$. The overall density of the network $\lambda=\lambda_p\bar{c}$ is fixed. The relative gain of MIMO IA compared to non-cooperative SISO is increasing for increasing number of cluster points. If we focus on the case of seven cluster points, the probability of success for $d_{ii}=[0.5,...,1]$ is increased by more than a factor of two. Therefore, local cooperation significantly increases the performance of the system whenever the signals of many strong interfering nodes can be aligned.

\begin{figure}[h]
  \centering
  \includegraphics[width=7.5cm]{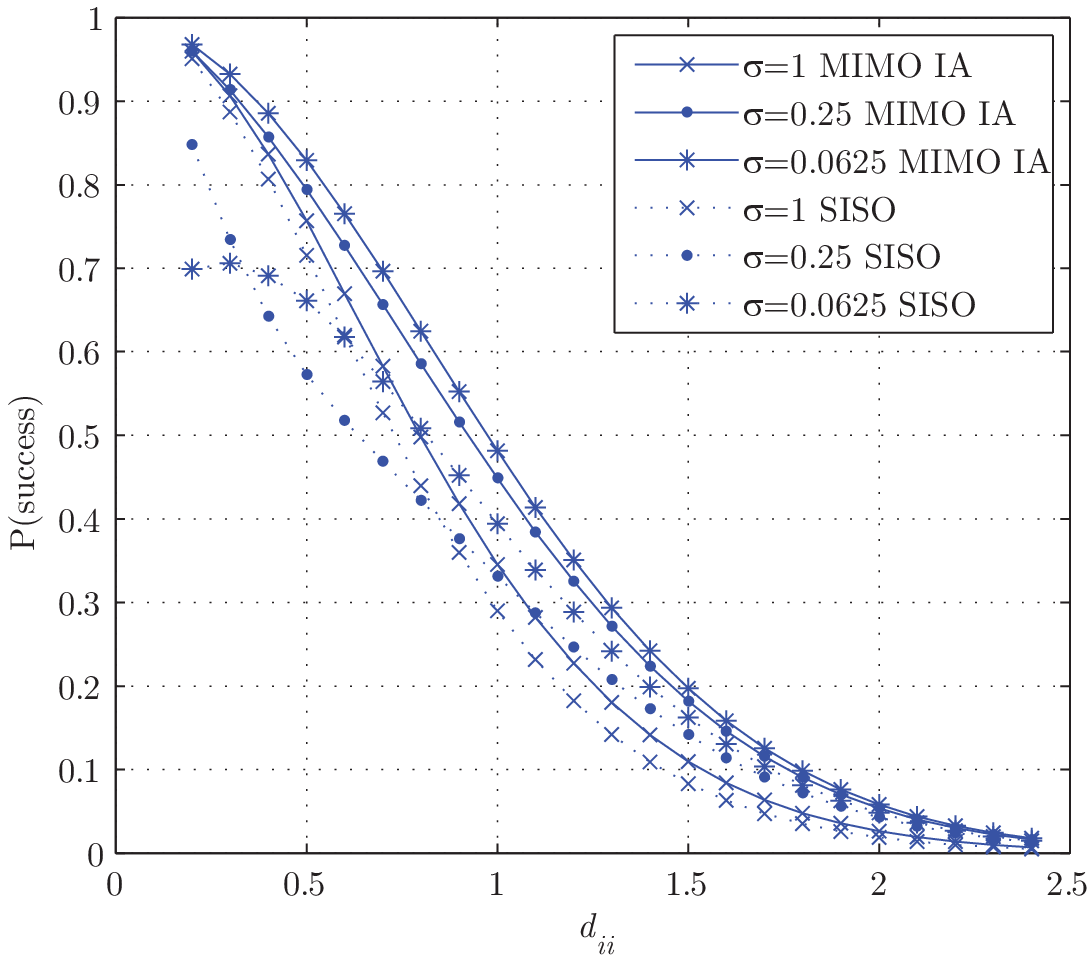}
  \caption{$\mathbb{P}(\textrm{success})$ versus $d_{ii}$ for $\lambda_p=0.25$, $\bar{c}=3$, $T=0.1$, and $\alpha=4$, comparison of SISO clustered network with MIMO intra-cluster IA}
  \label{Fig:Res2}
\end{figure}

In Fig. \ref{Fig:Res2}, we measure the performance gain of the MIMO intra-cluster IA scheme compared to the corresponding SISO network with the same network density for varying cluster spread. The density of the parent process is $\lambda_p=0.25$ and three cluster points $\bar{c}=3$ are considered. The smaller the scattering parameter $\sigma$ the better the IA strategy performs in terms of probability of success over the whole range of $d_{ii}$. However, for the non-cooperative SISO settings that suffer from intra- and inter-cluster interference, the performance is not monotonous with $\sigma$. E.g. the SISO network with $\sigma=0.0625$ outperforms the other SISO settings only in the range $d_{ii}>0.6$. The intra-cluster interference dominates the performance for small distances, i.e. $d_{ii}<0.6$. Here, the use of multiple antennas at each node, coupled with intra-cluster IA can significantly increase the performance. For $\sigma=0.25$ a maximum relative gain of 40\% is achieved at $d_{ii}=0.5$. If the scattering of the cluster points becomes larger (see $\sigma=1$ in Fig. \ref{Fig:Res2}), the benefit of intra-cluster interference suppression decreases.

\begin{figure}[h]
  \centering
  \includegraphics[width=7.5cm]{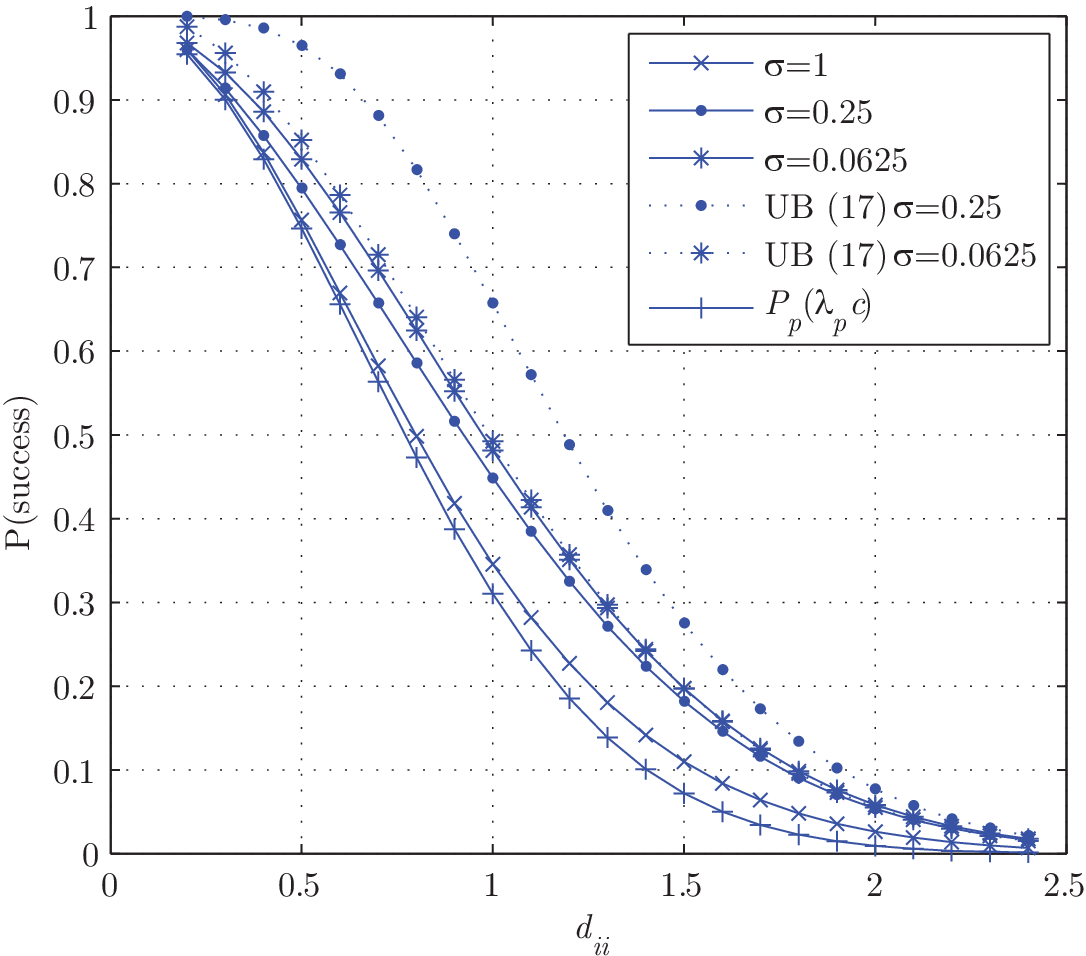}
  \caption{$\mathbb{P}(\textrm{success})$ versus $d_{ii}$ for $\lambda_p=0.25$, $\bar{c}=3$, $T=0.1$, and $\alpha=4$, closed-form upper bound and comparison with $P_p(\lambda_p \bar{c})$}
  \label{Fig:Res1}
\end{figure}

In Fig. \ref{Fig:Res1} the closed-form upper bound (\ref{equ_UB_closed_form}) for the same network setting as in Fig. \ref{Fig:Res2} is plotted.
For $\sigma=0.0625$ the upper bound is tight for $d_{ii}>1$. We also compare the probability of success of the intra-cluster IA scheme with the performance of a SISO network whose underlying node distribution is a corresponding homogeneous Poisson point process with intensity $\lambda=\lambda_p\bar{c}$. The probability of success for that case $P_p(\lambda_p\bar{c})=\exp(-\lambda_p\bar{c} d_{ii}^2T^{2/\alpha}C(\alpha))$ with $C(\alpha)=2\pi^2/\alpha\csc(2\pi/\alpha)$ \cite{GantiHaenggiit09} is also plotted in Fig. \ref{Fig:Res1}. It performs only slightly worse than the MIMO IA for $\sigma=1$. Increasing the spreading of the cluster points increases the overall spatial randomness in the network. Therefore, we conjecture that $P_p(\lambda_p\bar{c})$ is the limiting $\mathbb{P}(\textrm{success})$ for $\sigma\rightarrow\infty$.

\section{Conclusion}

The performance in terms of outage probability of interference alignment among a finite number of users was evaluated in a large clustered wireless ad hoc network. This strategy was shown to significantly increase the probability of successful transmission compared to non-cooperative networks. The gain stems from the suppression of dominant sources of interference in the vicinity of the intended transmitter and comes with the cost of deploying multiple antennas per node. In the light of those results, the proposed interference alignment scheme would be beneficially applied for networks with small cluster density and small clusters that contain many nodes.


\section*{Acknowledgment}
\small This work was supported in parts by the Vienna Science and Technology Fund (WWTF) through the project COCOMINT of the Vienna Telecommunications Research Center (FTW), by the COMET competence center program of the Austrian government and the City of Vienna through the FTW I0 project, and inspired by our activity in the European Commission FP7 Newcom++ network of excellence.



%

\end{document}